\documentclass{emulateapj} 
 
\slugcomment{Accepted for Publication; To appear in the ApJS Special
Issue on Spitzer, September 2004}

\shorttitle{Confusion in the infrared: {\it Spitzer} and beyond} 
\shortauthors{H. Dole et al.} 
 
\begin{document} 
 
 
\title{Confusion of Extragalactic Sources in the Mid- and 
Far-Infrared: \\ 
{\it Spitzer} and beyond} 
 
 
\author{H. \ Dole\altaffilmark{1,2}, 
G. H.\ Rieke\altaffilmark{1}, 
G. \ Lagache\altaffilmark{2}, 
J-L. \ Puget\altaffilmark{2}, 
A. \ Alonso-Herrero\altaffilmark{1}, 
L. \ Bai\altaffilmark{1}, 
M. \ Blaylock\altaffilmark{1}, 
E. \ Egami\altaffilmark{1}, 
C. W.\ Engelbracht\altaffilmark{1}, 
K. D.\ Gordon\altaffilmark{1}, 
D. C.\ Hines\altaffilmark{1,3}, 
D. M. \ Kelly\altaffilmark{1}, 
E. \ Le Floc'h\altaffilmark{1}, 
K. A.\ Misselt\altaffilmark{1}, 
J. E.\ Morrison\altaffilmark{1}, 
J. \ Muzerolle\altaffilmark{1}, 
C. \ Papovich\altaffilmark{1}, 
P. G. \ P\'erez-Gonz\'alez\altaffilmark{1}, 
M. J. \ Rieke\altaffilmark{1} 
J. R.\ Rigby\altaffilmark{1}, 
G. \ Neugebauer\altaffilmark{1}, 
J. A. \ Stansberry\altaffilmark{1}, 
K. Y. L. \ Su\altaffilmark{1}, 
E. T.\ Young\altaffilmark{1}, 
C. A. \ Beichman\altaffilmark{4}, 
P. L. \ Richards\altaffilmark{5} 
} 
\altaffiltext{1} {Steward Observatory, University of Arizona, 933 N 
Cherry Ave, Tucson, AZ 85721, USA} 
\altaffiltext{2} {Institut d'Astrophysique Spatiale, b\^at 121, 
Universit\'e Paris-Sud, F-91405 Orsay Cedex} 
\altaffiltext{3} {Space Science Institute 
4750 Walnut Street, Suite 205, Boulder, Colorado 80301} 
\altaffiltext{4} {Michelson Science Center, CalTech, Pasadena, CA, USA} 
\altaffiltext{5} {Dept. of Physics, 345 Birge Hall, Berkeley, CA, USA} 
 

 
\begin{abstract} 
We use the source counts measured with the Multiband Imaging 
Photometer for {\it Spitzer} (MIPS) at 24, 70, and 160~$\mu$m to
determine the 5-$\sigma$ confusion limits due to extragalactic sources: 
$56 \mu$Jy, 3.2 and 40 mJy at 24, 70 and 160~$\mu$m, respectively. 
We also make predictions for confusion limits for a number of proposed
far infrared missions of larger aperture (3.5 to 10m diameter). 
\end{abstract} 
 
 
 
 
%
\section{Introduction} 
 
In addition to detector/photon noise, cosmological surveys in the
far-infrared (FIR) spectral range are limited in depth by: 1.)
structure in the infrared cirrus emission; 
and 2.) confusion by extragalactic sources. 
The first of these limitations can be avoided for some programs by
observing in particularly low-background regions on the sky
The second limitation arises because the high density of faint (resolved or
unresolved) distant galaxies creates signal fluctuations in the
telescope beam (Condon 1974; Franceschini et al. 1989; Helou \& Beichman 1990; 
Rieke et al. 1995; Dole et al. 2003; Takeuchi \& Ishii 2004, for instance).  
Because distant galaxies are distributed roughly isotropically and 
with a high density compared to the beam size, this noise is unavoidable. 
 
Extragalactic confusion noise can be robustly estimated by 
measurements of source counts combined with modeling to extend the 
counts to faint levels. We use new determinations of number 
counts in the three Multiband Imaging Photometer for {\it Spitzer} 
(MIPS, Rieke et al., 2004) bands, 24, 70, and 160~$\mu$m 
\cite[]{dole2004, papovich2004}, and a model fitting all those 
observables \cite[]{lagache2004} to determine more accurate limits for
extragalactic confusion than have been available previously.  
Extragalactic confusion noise does not strictly follow Gaussian
statistics. Therefore, we discuss confusion limits in four different
ways that are appropriate to various measurement situations: the
photometric criterion and source density criterion (hereafter SDC)
\cite[]{dole2003}, and the levels deduced from the source densities of
one source per 20 and 40 independent beams. We parameterize the noise
as a "$5-\sigma$" limit calculated as if it were Gaussian, because it
is difficult to derive any other simple metric. All the definitions
and values relative to MIPS {\it Spitzer} beams are summarized in
Table~1 of \cite{dole2003}. 
 
We summarize the confusion limits for {\it Spitzer} in its three 
far-infrared bands in Table~\ref{tab:conf}. 
The situation is different at 24 and 70~$\mu$m than at 160~$\mu$m. In
the two first bands, where the background is resolved to a significant
extent, the confusion mainly results from the high density of resolved
sources and their interference with extraction of fainter ones: the
SDC is the appropriate measure (and the classical photometric
criterion underestimates the confusion level). In the third band,
where the background is not well resolved, the confusion results from a 
population fainter than the sensitivity limit. In the latter case, 
confusion (and CIB fluctuation) properties are directly linked to 
galaxy populations not directly detectable but which modulate
the background level: the photometric criterion is appropriate.

%
\section{Confusion in the Mid- and Far- Infrared} 
\label{sect:conf} 
%
\subsection{Confusion of Extragalactic Sources at 24~$\mu$m} 
\label{sect:mips024} 
The available measurements extend well into the extragalactic
confusion regime at 24$\mu$m, and the detector performance is also
well understood even for long integrations. Therefore, we use this
band to develop the general principles applicable to determining the
confusion limits in {\it Spitzer} 
mid- and far-infrared imaging data. 
\subsubsection{Confusion Limit Calculation} 
 
Our confusion estimates are based on the methodology described by
\cite{dole2003}. We have used the number counts determined by
\cite{papovich2004}, extrapolated to fainter flux limits according
to the model of \cite{lagache2004}. Because these counts indicate
that the background will be largely resolved into individual sources,
the appropriate measure of the confusion is the SDC. We obtain
56~$\mu$Jy for the 5-$\sigma$ confusion level, corresponding to 12
beams per source. It appears that this confusion level is in perfect
agreement with the $5\sigma$ pre-launch predictions of \cite{xu2001},
even if it was derived differently.
If it were limited by photon noise only, the instrument would reach a
detection limit of 56~$\mu$Jy 5-$\sigma$ in 1900 seconds of
integration \cite[]{rieke2004}, so the model predicts that the gain in
signal to noise ratio will have leveled out significantly for
integrations of this length. 

%
\begin{figure} 
\plotone{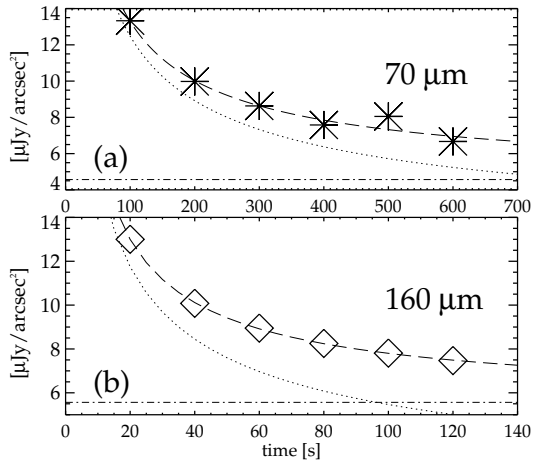} 
\caption{ 
Evolution of $\sigma_{tot}$ (resulting contribution from the confusion
noise and instrument noise, and derived from the Gaussian fit in the
brightness map pixel histogram) as a function of integration time,
with a fit (dash) of the form:  
$\sigma_{tot}^2 = \sigma_{inst}^2 + \sigma_{conf Br}^2 = At^{-1} + C^2$. 
Dot-dash: constant term $C$.
Dot: $\sqrt{A/t}$ term.  
Top panel (a): 70~$\mu$m. 
Middle panel (b): 160~$\mu$m. 
Notice the different scales in time (seconds) and $\sigma_{tot}$ 
(in brightness $\mu$Jy/arcsec$^2$).
\label{fig:letter_confusion_rms}} 
\end{figure} 

There is excellent agreement between the observed 80\% completeness
level and source density of \cite{papovich2004}, and our SDC confusion
level. However, it should be possible in principle to integrate below
the 56~$\mu$Jy level, on a {\it selected field} of very low source
density. In the "GOODS Test Field" in ELAIS N1 (described in Papovich
et al. 2004), we estimate the area suitable for a deeper integration
to be about 5\% of the field area. 

\subsubsection{Noise Analysis}

We desired a test of these predictions that was as much as possible
independent of assumptions about the infrared galaxy population. For
this purpose, we have characterized the noise in the  24~$\mu$m data
from the ELAIS N1 field, the deepest observation obtained to date at
this wavelength. We selected a very cleanly reduced region in the
field, about 2 x 4 arcmin in size. We prepared two versions of the
image in this region, both reduced identically, but one with an
integration of 630 seconds and the other with an integration of 3800
seconds. We determined the pixel signal histogram in two ways.
1.) On a small region that also appeared to be free of detected
sources, we verified that the standard deviation as measured in these
histograms scaled inversely with the square root of the integration
time.
2.) On the entire 2 x 4 arcmin region, we fitted it with a Gaussian of
width fixed to the expectation for detector/photon noise. We required
this Gaussian to fit only the negative side of the histogram, on the
assumption that there were no negative sources. We took the departure
of the measured histogram from this fit toward positive fluctuations
to be the influence of (at least) sources in the field. We measured
the extension of the distribution toward positive values at half
maximum. We found that the positive-going width of the distribution
was larger than the pure detector/photon noise expectation by a factor
of 1.7, in qualitative agreement with the effects of confusion.
These excess fluctuations likely result from a combined effect
of extragalactic sources, faint cirrus and zodiacal light gradient. It
is not clear at this stage which component dominates the fluctuations.

%
\begin{deluxetable}{lccc} 
\tablewidth{0pt} 
\tablecaption{{\it Spitzer} MIPS Confusion Levels$^a$ \label{tab:conf}} 
\tablehead{ 
\colhead{} & 
\colhead{24~$\mu$m} & 
\colhead{70~$\mu$m} & 
\colhead{160~$\mu$m} \\ 
\colhead{} & 
\colhead{[$\mu$Jy]} & 
\colhead{[mJy]} & 
\colhead{[mJy]} \\ 
} 
\startdata 
SDC$^b$ & 56    & 3.2  &  40 \\ 
20b$^c$ & 71    & 3.5  &  45 \\ 
40b$^d$ & 141   & 6.3  &  63 \\ 
Phot$^e$ & 8   & 0.7  &  45 \\ 
\enddata 
\tablecomments{ 
(a) with \cite{lagache2004} model.
(b) Using the Source Density criterion \cite[]{dole2003}. 
(c) Using the flux corresponding to one source per 20 
beams.
(d) Using the flux corresponding to one source per 40 
beams.
(e) Using the standard photometric criterion and $q=4$, for
illustration.
} 
\end{deluxetable} 

\subsubsection{Monte Carlo Simulation}

To empirically quantify the effect of confusion, we carried out a
Monte Carlo simulation of source extraction under the conditions
appropriate for the {\it Spitzer} deep 24$\mu$m exposures.
The approach is described in detail by Rieke et al. (1995). We built
up a test field by distributing confusing sources randomly according
to a power law distribution matching the faint {\it Spitzer} number
counts. Each source was entered as an Airy pattern. A test source of
known amplitude was added to the center of the array, along with
Gaussian noise. The sources were then identified using a modified
CLEAN algorithm and finally the signal to noise was measured in a
master array built up from the results of the CLEAN process, and in
extraction apertures of various sizes. An important aspect of this
simulation is that it combines the effects of neighboring bright
sources and of the underlying, unresolved distribution of faint ones,
in a consistent manner. It should give a good measure of the confusion
noise independent of the division between source density and
photometric criteria. 

In the simulation, we excluded all objects brighter than 400$\mu$Jy to
avoid undo noise from bright-source artifacts. The first set of runs
tested the extraction of a 56$\mu$Jy source in an 0.8 $\lambda$/D
beam, the beam size previously indicated to provide optimum
performance in a heavily confusion-limited situation (Rieke et
al. 1995 - this result was confirmed by the new calculations). We made
1200 runs for an integration time long enough to drive detector/photon
noise down to 12.5$\mu$Jy, 5-$\sigma$. They yielded a net 5-$\sigma$
limit of 60$\mu$Jy: removing the detector/photon noise leaves
59$\mu$Jy of confusion noise. That is, this approach agrees well with
the SDC-determined limit of 56$\mu$Jy. 

We also simulated the results to be expected from shorter integration
times. For example, if the 5-$\sigma$ detector/photon noise limit was
set to 65$\mu$Jy, then the indicated 5-$\sigma$ level of confusion
noise was 76$\mu$Jy, significantly poorer than from the simulation of
very long integrations. This effect probably results from the
increased uncertainty in source centroiding and the resulting lower
accuracy in extracting accurate source measurements from a confused
field. To test this hypothesis further, we simulated extraction of a
36$\mu$Jy source in the high signal-to-noise integration case, and
found that the indicated 5-$\sigma$ confusion limit rose to 64$\mu$Jy,
confirming the effect.

\subsection{Confusion by Extragalactic Sources at 70~$\mu$m} 
\label{sect:mips070} 
 
At 70~$\mu$m, we again use the number counts \cite[]{dole2004} as the
basic input to determining the confusion level. The updated model of 
\cite{lagache2004} was used to extrapolate the counts and to derive
updated confusion limits. The use of a model is critical in this case
because the contribution of unresolved sources is not negligible. We
derive a confusion level at 70~$\mu$m of 3.2~mJy using the SDC
(Table~\ref{tab:conf}). The differential source counts are almost
flat (when divided by the Euclidean component), and the contribution
from unresolved sources is much smaller than that of the resolved sources. 
These results demonstrate that the SDC estimate is the appropriate one,
that is, the confusion is dominated by faint resolved sources rather
than the unresolved background due to even fainter objects. Further
details are given in Table~\ref{tab:conf}. From the instrument
radiometric model, we estimate that about 1800 seconds of integration
would be required to reach this limit.

Again, we sought to check these results by a pure fluctuation analysis
on the data without referring to galaxy population models. We used the
data described by \cite{dole2004} for the Chandra Deep Field South.
We determined the evolution of $\sigma_{tot}$, the standard deviation
of a Gaussian fitted to the surface brightness distribution as a check
of the results from extrapolating number counts downward. Data 
were combined into 6 mosaics corresponding to 100s to 600s integration 
time per sky pixel with 100s steps. Figure~\ref{fig:letter_confusion_rms}a 
shows the evolution of $\sigma_{tot70}$ with time.  
We do not observe substantial flattening in the $\sigma_{tot70}$ time
evolution. We conclude that MIPS 70~$\mu$m surveys do not reach yet
the confusion limit after 600s of integration. An estimate of the 
confusion level is given by fitting the time evolution of 
$\sigma_{tot70}$. We find that the detector/photon noise will
be roughly equal to the confusion noise at $\ge$ 800 seconds of
integration, with large uncertainties because the fluctuation curve is
still dropping almost as the square root of the integration time at
the longest integration available. As at 24$\mu$m, this result is in
satisfactory agreement with the integration time predicted by the SDC
modeling.  
 
%
\subsection{Confusion by Extragalactic Sources at 160~$\mu$m} 
\label{sect:mips160} 
 
The data used at 160~$\mu$m are also described by \cite{dole2004}.
The \cite{lagache2004} model predicts a confusion level of 40 mJy
(Table~\ref{tab:conf}).
From the instrument radiometric model, we estimate that about 70
seconds of integration would be required to reduce the instrument and
photon noise to the level of the confusion noise.
 
A similar fluctuation analysis as at 70~$\mu$m was conducted at 
160~$\mu$m, where 6 mosaics corresponding to integration times of 20 
to 120s (with 20s steps) were studied. 
Analyzing the fluctuations is more difficult in this case since bright 
sources in the Euclidean regime contaminate the statistics, and since 
the map S/N is not uniform. Nevertheless, we estimate from
Figure~\ref{fig:letter_confusion_rms}b that the confusion and
detector/photon noise should be equal at about 95 seconds of
integration, in good agreement with the result from the SDC analysis.

%
\begin{figure} 
\plotone{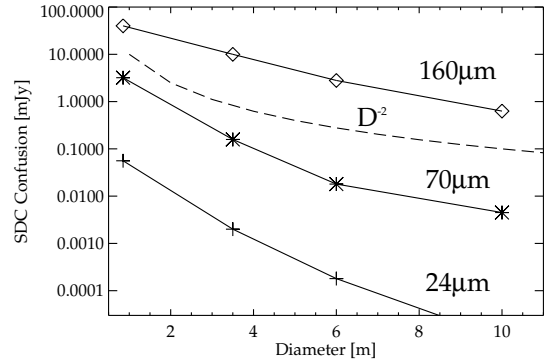} 
\caption{Confusion level vs Telescope Diameter, predicted by the 
Source Density Criterion \cite[]{dole2003} with the updated model of 
\cite{lagache2004}, at 24 (plus), 70 (star) and 160~$\mu$m (diamond). 
Diameters refer to {\it Spitzer}, Herschel and SPICA, JWST and SAFIR. 
Dash: inverse square diameter law shown for illustration. 
\label{fig:letter_confusion_conf3wave}} 
\end{figure} 

%
\subsection{Confusion by Galactic Cirrus} 
\label{sect:cirrus} 

Another sensitivity limitation arises due to the structure of the IR
cirrus. To estimate how this cirrus emission may affect the source
detectability, we compared the 80\% completeness limit in sky regions
characterized by different cirrus background levels, using simulations
as described in \cite{papovich2004}. We used a dedicated engineering
observation in Draco of a {\it bright cirrus}, of HI column density
$n_{HI}$ varying between $4$ and $14 \times 10^{20} cm^{-2}$.
At 24$\mu$m, we find a relatively weak effect, and derive a
completeness degradation of 15\% ($\sim 50 \mu$Jy increase from
$340\mu$Jy) between the dark and the bright parts of the cirrus field.
The effects of cirrus are more conspicuous at 70$\mu$m.
We reach a 80\% completeness in Draco of $\sim$17mJy and $\sim$27mJy.
In a low-cirrus field (e.g., Marano) and for a similar integration
time (100s), this level drops at $\sim$12mJy.
We compared the estimates in Draco with those provided by the
Performance Estimation Tool of the {\it Spitzer\,} Science Center and
found that the measured value variations as a function of the cirrus
strength are in general agreement (within 30\%) with those estimated
by the tool from low to medium background. This comparison will be
refined as we continue to acquire far-infrared data.

%
\section{Implications for Future Observatories} 
\label{sect:discussion} 
 
A number of cryogenically-cooled space telescopes have been proposed 
for the MIR, the FIR and the submillimeter spectral 
ranges. Table~\ref{tab:conf2} summarizes the main characteristics of 
some of these observatories.  
Herschel \cite[]{pilbratt2001},
JWST \cite[]{gardner2003}, 
SPICA \cite[]{matsumoto2003} and 
SAFIR \cite[]{yorke2002}, 
have at least one photometric channel in common with MIPS.  
As examples, we focus on Herschel-PACS at 75 and 170~$\mu$m, 
on JWST-MIRI at 24~$\mu$m, and on SPICA and SAFIR at 
24, 70 and 160~$\mu$m, assuming in each case that the MIPS filters 
will be used. 

For each of these observatories, we compute predictions for the 
confusion level for unbiased surveys using the \cite{lagache2004} 
model of source counts.  
We assume a Gaussian beam profile for these future observatories, with 
a FWHM of $1.22 \lambda / D$, $\lambda$ being the wavelength and $D$ the 
diameter of the primary telescope mirror, given in Table~\ref{tab:conf2}. 
The underlying assumption for the deepest surveys to be made by these 
planned facilities, is that they will be confusion-limited. This means 
we did not take into account other sources of noise, for instance 
photon noise due to insufficient integration times, or thermal 
background due to the warm telescope -- by design, Herschel and JWST 
might be in the latter case. 
Normally background limited photon noise observations would give a 
sensitivity limit scaling as aperture squared for a diffraction 
limited system. Figure~\ref{fig:letter_confusion_conf3wave} shows that 
confusion noise at 24 and 70~$\mu$m drops much faster than as aperture 
squared (dash) because source counts are shallower below fluxes where
most of the CIB has been resolved into sources.
That is why the next generation of large far infrared telescopes will
be much less confusion limited than {\it Spitzer}. 
 
In Table~\ref{tab:resol_cib}, we use the confusion level given by the 
SDC, and compute the fraction of the CIB potentially resolved into 
sources.  
In the MIR, a significant step will be made with the 4m-class space 
telescope: as an example, SPICA would potentially resolve 98\% of the 
CIB at 24~$\mu$m. All ($>99$\%) of the CIB would be resolved with JWST 
or SAFIR (although doing so with JWST would require extremely long 
integrations). 
In the FIR, Herschel would resolve a significant fraction of the CIB at 
70 and 160~$\mu$m (resp. 93 and 58\%, again with extremely long 
integrations). SAFIR will ultimately nearly resolve all of it ($>94$\%).

%
\begin{deluxetable}{lccc} 
\tablewidth{0pt} 
\tablecaption{Telescopes and Predicted Confusion Levels \label{tab:conf2}} 
\tablehead{ 
\colhead{} & 
\colhead{Herschel$^a$} & 
\colhead{JWST$^b$} & 
\colhead{SAFIR} \\ 
\colhead{} & 
\colhead{SPICA} & 
\colhead{} & 
\colhead{} 
} 
\startdata 
Diameter (m)  & 3.5 & 6.0 & 10.0 \\ 
24~$\mu m$ SDC$^c$ [$\mu$Jy]&  2  & 0.18 & $<$0.01$^f$\\ 
70~$\mu m$ SDC$^c$ [mJy]    & 0.16& --  & 0.004\\ 
160~$\mu m$ SDC$^d$ [mJy]   & 10  & --  & 0.6\\ 
\enddata 
\tablecomments{ 
(a) With PACS.
(b) With MIRI.
(c) cf notes a and b in Table~\ref{tab:conf}.
(d) outside the range of the current model flux grid. 
} 
\end{deluxetable}

%
\begin{deluxetable}{lccc} 
\tablewidth{0pt} 
\tablecaption{Potential Resolution of the Cosmic Infrared Background \label{tab:resol_cib}} 
\tablehead{ 
\colhead{} & 
\colhead{24 $\mu$m$^a$} & 
\colhead{70 $\mu$m$^a$} & 
\colhead{160 $\mu$m$^a$} 
} 
\startdata 
{\it Spitzer} & 74\% & 59\% & 18\% \\ 
{Herschel$^b$/SPICA} & 98\% & 93\% & 58\% \\ 
{JWST$^b$} & 99\% & -- & -- \\ 
{SAFIR} & 100\% & 99\% & 94\% \\ 
\enddata 
\tablecomments{ 
(a) Using the CIB value from \cite{lagache2004} and 
using the limiting flux using the SDC limit and assuming 
confusion-limited surveys.
(b) This hypothesis might not be valid for Herschel and JWST. 
} 
\end{deluxetable} 

%
\section{Conclusions} 
\label{sect:conclusion} 
Using MIPS {\it Spitzer} data at 24, 70 and 160~$\mu$m, the source 
density measured by \cite{papovich2004} and \cite{dole2004} together 
with the modeling of \cite{lagache2004} has allowed us to derive the 
confusion limits for {\it Spitzer} in the mid to far infrared.  
We tested the model results with a Monte Carlo simulation at
24$\mu$m and with a fluctuations analysis at all 3 wavelengths.
The agreement is uniformly very good. 

At 24 and 70~$\mu$m, confusion is mostly due to the high density of
resolved sources, and at 160~$\mu$m, confusion is mainly due to faint
unresolved sources. Studying the FIR fluctuations at this wavelength
is thus a tool to constrain the nature of the faint galaxies, beyond
the confusion limit. 
 
We also derive confusion limits for future space IR observatories.
We show that future large-aperture
missions will gain in confusion-limited sensitivity substantially
faster than as aperture squared for wavelengths $\le$ 100$\mu$m,
allowing them to reach very deep detection limits. For example, the
CIB should be fully resolved into sources in the MIR and FIR with
SAFIR observations.  
 
%
\acknowledgments 
This work is based on observations made with the {\it Spitzer}
Observatory, which is operated by the Jet Propulsion Laboratory, 
California Institute of Technology under NASA contract 1407. 
We thank the funding from the MIPS project, which is supported by 
NASA through the Jet Propulsion Laboratory, subcontract \#960785. 
We warmly thank J. Cadien.

%


\begin{thebibliography}{}  
 
\bibitem[{Condon}(1974)]{condon74} 
{Condon}, J.~J. 
\newblock 1974, {\em \apj}, 188:279. 
 
\bibitem[{Dole} et~al.(2003)]{dole2003} 
{Dole}, H, {Lagache}, G, \& {Puget}, J.~L. 
\newblock 2003, {\em \apj}, 585:617. 
 
\bibitem[{Dole} et~al.(2004)]{dole2004} 
{Dole}, H, {Le Floc'h}, E, {P\'erez-Gonz\'alez}, P. G., et~al. 
\newblock 2004, {\em \apjs}, this volume. 
 
\bibitem[{Franceschini} et~al.(1989)]{franceschini89} 
{Franceschini}, A, {Toffolatti}, L, {Danese}, L, \& {De Zotti}, G. 
\newblock 1989, {\em \apj}, 344:35. 
 
\bibitem[{Gardner}(2003)]{gardner2003} 
{Gardner}, J.~P. 
\newblock 2003, IAU, Joint Discussion 8, page~10 
 
\bibitem[{Helou} \& {Beichman}(1990)]{helou90} 
{Helou}, G, \& {Beichman}, C.~A. 
\newblock 1990, ESA SP-314, p117  
 
\bibitem[{Lagache} et~al.(2004)]{lagache2004} 
{Lagache}, G, {Dole}, H, \& {Puget}, J.~L., et~al. 
\newblock 2004, {\em \apjs}, this volume. 
 
\bibitem[{Lonsdale} et~al.(2003)]{lonsdale2003} 
{Lonsdale}, C.~J, {Smith}, H.~E, {Rowan-Robinson}, M, et~al. 
\newblock 2003, {\em \pasp}, 115:L897. 
 
\bibitem[{Matsumoto}(2003)]{matsumoto2003} 
{Matsumoto}, T. 
\newblock 2003, SPIE, 4850:1091 
 
\bibitem[{Papovich} et~al.(2004)]{papovich2004} 
{Papovich}, C, {Dole}, H, {Egami}, E, et~al. 
\newblock 2004, {\em \apjs}, this volume. 
 
\bibitem[{Pilbratt}(2001)]{pilbratt2001} 
{Pilbratt}, G.~L. 
\newblock 2001, ESA-SP 460:13 
 
\bibitem[{Rieke} et~al.(1995)]{rieke95} 
{Rieke}, G.~H, {Young}, E.~T, \& {Gautier}, T.~N. 
\newblock 1995, {\em \ssr}, 74:17. 
 
\bibitem[{Rieke} et~al.(2004)]{rieke2004} 
{Rieke}, G.~H, {Young}, E.~T, {Engelbracht}, C, et~al. 
\newblock 2004, {\em \apjs}, this volume. 
 
\bibitem[{Takeuchi} \& {Ishii}(2004)]{takeuchi2004} 
{Takeuchi}, T.~T \& {Ishii}, T.~T. 
\newblock 2004, {\em \apj}, 604:40. 
 
\bibitem[{Xu} et~al.(2001)]{xu2001} 
{Xu}, C.~K, {Lonsdale}, C.~J, {Shupe}, D.~L, O'Linger, J., \& Masci, F. 
\newblock 2001, {\em \apj}, 562:179. 
 
\bibitem[{Yorke} et~al.(2002)]{yorke2002} 
{Yorke}, H.~W, {Bock}, J.~J, {Dragovan}, M.~W et~al. 
\newblock 2002, {\em AAS}, 2011:5104. 
 
\end{thebibliography}
\end{document}